\documentclass[a4paper, amsfonts, amssymb, amsmath, reprint, showkeys, nofootinbib, twoside]{revtex4-1}
\usepackage[english]{babel}
\usepackage[toc,page]{appendix}
\usepackage{float}
\usepackage{xcolor}
\usepackage{subcaption}
\usepackage{dcolumn}
\usepackage{bm}
\usepackage{multirow}
\usepackage{array}
\usepackage{siunitx} 
\usepackage{booktabs}
\usepackage{calligra}
\usepackage{colortbl}
\usepackage{graphicx}
\usepackage{mathtools}
\usepackage{amsmath}
\usepackage{mathpazo}
\usepackage[export]{adjustbox}
\usepackage{mathrsfs}
\usepackage{natbib}
\usepackage[font=footnotesize,labelfont=bf, justification=justified,
   format=plain]{caption}
\DeclareSymbolFont{pxlettersA}{U}{pxmia}{m}{it}
\SetSymbolFont{pxlettersA}{bold}{U}{pxmia}{bx}{it}
\DeclareMathSymbol{\varg}{\mathord}{pxlettersA}{49}
\usepackage{xcolor,colortbl}

\usepackage{calligra} 

\DeclareMathAlphabet{\mathcalligra}{T1}{calligra}{m}{n}
\DeclareFontShape{T1}{calligra}{m}{n}{<->s*[2.2]callig15}{}

\definecolor{Gray}{gray}{0.85}
\definecolor{Orange}{rgb}{1,0.7,0}
\definecolor{Purple}{rgb}{0.25,0,0.78}
\definecolor{White}{rgb}{1.0,1.0,1.0}

\newcolumntype{a}{>{\columncolor{Gray}}c}
\newcolumntype{b}{>{\columncolor{White}}c}

\usepackage[utf8]{inputenc}
\usepackage[colorinlistoftodos, color=green!40, prependcaption]{todonotes}
\usepackage{amsthm}
\usepackage{mathtools}
\usepackage{physics}
\usepackage{xcolor}
\usepackage{graphicx}
\usepackage[left=23mm,right=13mm,top=35mm,columnsep=15pt]{geometry} 
\usepackage{adjustbox}
\usepackage{placeins}
\usepackage[T1]{fontenc}
\usepackage{lipsum}
\usepackage{csquotes}

\begin{document}
\title{Stochastic consensus dynamics for decentralized decision systems}

    \author{Andr\'e L. M. Vilela}
    \affiliation{F\'isica de Materiais, Universidade de Pernambuco, Recife, PE 50720-001, Brazil}
    \affiliation{Departamento de F\'isica, Universidade Federal de Pernambuco, Recife, PE 50670-901, Brazil}
    
    \author{Caio B. L. Silva}
    \affiliation{F\'isica de Materiais, Universidade de Pernambuco, Recife, PE 50720-001, Brazil}
    
    \author{Kenric P. Nelson}
    \affiliation{Photrek, Inc., Watertown, MA 02472, USA}
    
    \author{Emilio Cobanera}
    \affiliation{Department of Physics, SUNY Polytechnic Institute, Utica, NY 13502, USA}
    \affiliation{Department of Physics and Astronomy, Dartmouth, Hanover, NH 03755, USA}
    
    \author{Gaogao Dong}
    \affiliation{School of Mathematical Sciences, Jiangsu University, Zhenjiang, 212013, Jiangsu, China}

\date{\today} 

\begin{abstract}
A fundamental mechanism underlying collective phenomena in social, technological, and economic systems is decentralized decision-making, in which the behavior of individual agents follows from local interactions in the absence of central coordination. Consensus formation is a central feature of such systems, with relevance to blockchain networks, distributed artificial intelligence, autonomous multi-agent systems, distributed control, and collective decision networks. We investigate consensus formation using a stochastic consensus model on random networks and examine how initial conditions, network connectivity, and system size shape the emergence of unanimous states. We show that the stochastic dynamics strongly amplify small initial majorities, progressively suppress the competing state, and drive the system toward a predictable collective outcome. Network connectivity primarily controls the efficiency of this process: increasing connectivity accelerates the propagation of local agreement and reduces the likelihood that fluctuations reverse the initially dominant state, although these gains gradually saturate in highly connected networks. System size produces a complementary effect. Larger networks require more individual updates to reach unanimity, but they are also increasingly reliable in selecting the state favored by the initial majority. Finite-size analysis shows that the range of initial conditions associated with uncertain outcomes becomes progressively narrower as the network grows, decreasing approximately with the inverse square root of the system size. These results reveal a collective amplification mechanism by which weak initial asymmetries become increasingly decisive in large decentralized networks, and they provide a simple framework for understanding the efficiency, predictability, and reliability of consensus formation in distributed decision systems.

\end{abstract}

\keywords{Consensus dynamics, Decentralized decision-making, Decentralized systems, Blockchain, Complex systems.}

\maketitle

\section{INTRODUCTION} \label{sec:intro}

Decentralized decision-making refers to the process by which a collection of autonomous agents, the system, reaches a collective decision through local interactions without intervention from a central authority or coordinator. In these systems, individual agents update their state based on partial information obtained from neighboring agents, and global agreement emerges from the accumulation of many locally-determined decisions --- that is, state updates. This mechanism for reaching consensus occurs in a wide range of natural, social, and technological systems. Examples include opinion formation in social networks \cite{vilela2018effect, katarzyna2005, galam2008, Shang2017, Piva2022, Goles2023, Fontanari2012, Castellano2009}, distributed artificial intelligence and large language models \cite{Dwork1988, baronchelli2025}, autonomous multi-agent systems \cite{Ma2017, Xie2021}, organizational and public governance \cite{Robertson2012, Boasson2023}, and blockchain infrastructures \cite{Mattos2020, Musilek2021, Fadda2022, Jayabalasamy2024, Schwartz2018, Li2020, Khacef2024, Alt2025}.

\begin{figure*}[ht]
    \centering
    \includegraphics[width=0.8\linewidth]{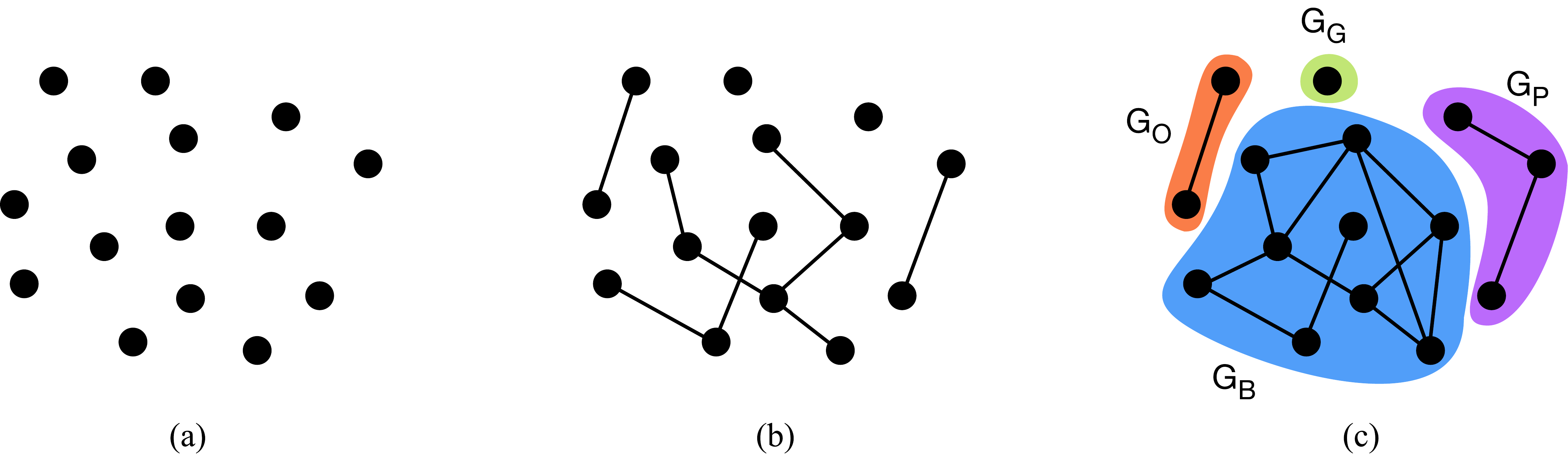}
    \caption{
    \textbf{Random network construction representation.} In (a), we illustrate $N = 15$ isolated nodes; in (b), the intermediate step of randomly selecting nodes and connecting them while avoiding double links. In (c), we show the completed process with $L = 15$ links, which yields a network with $\langle k \rangle = 2$. Note that this network has four components denoted by the colors green, orange, purple, and blue, each with $1$, $2$, $3$, and $9$ nodes, respectively.} 
    \label{fig:netvis}
\end{figure*}

A prominent technological realization of this paradigm is distributed ledger technology, which provides a decentralized infrastructure for recording, updating, and verifying information across a network of nodes. In these systems, a shared public ledger is maintained through consensus protocols, as exemplified by cryptocurrency platforms such as Bitcoin \cite{Alt2025, Nakamoto2008, Peck2017}.

The growing integration of decentralized systems with emerging technologies, including artificial intelligence, the Internet of Things, and large-scale digital infrastructures, has increased demand for scalable, reliable consensus mechanisms in modern interconnected environments. Maintaining a consistent collective state across distributed agents requires consensus dynamics capable of operating under heterogeneous network conditions, communication delays, node failures, and potentially conflicting or adversarial behavior \cite{Steve2016, Denis2018, Tiago2018, Thomas2019, Mishra2021, Hisham2023, Dragoni2020, Allende2023, Parida2023, Gharavi2024, Sinai2024, Chen2022}.

A central challenge in decentralized decision systems is determining whether small initial majorities can be reliably amplified into global agreement. This problem is closely related to the broader study of consensus dynamics on complex networks, where topology, local interactions, and stochastic fluctuations jointly shape the emergence of collective order \cite{Goles2023, Dwork1988, Ma2017, Xie2021, baronchelli2025, Piva2022}. Network science, together with voting and opinion-dynamics models, therefore provides a natural framework for investigating how interaction structure influences the speed, reliability, and scalability of consensus formation \cite{Kurths2021, granha2022opinion, de1992isotropic, Baronchelli2018, santos1995anisotropic, campos2003small, pereira2005majority, Anagnostakis2025}.

Several studies have examined the mechanisms that shape collective agreement in decentralized systems. Previous work has shown that voting anisotropy can weaken consensus under noisy conditions \cite{santos1995anisotropic}, that increasing connectivity and long-range interactions enhance the robustness of consensus formation \cite{pereira2005majority, campos2003small}, that central nodes in heterogeneous networks can gain a competitive advantage in consolidating consensus \cite{Fadda2022}, and that witnessing protocols can be analyzed through the frameworks of Shannon information entropy and Lyapunov stability \cite{Anagnostakis2025}. These studies demonstrate that consensus outcomes are strongly influenced by both network topology and local interaction rules. However, quantifying how efficiently decentralized systems amplify weak majorities into reliable collective decisions, and how the predictability of the final consensus scales with network connectivity and system size, remains an important challenge.
    
\begin{figure*}[ht]
    \centering
    \includegraphics[width=1.0\linewidth]{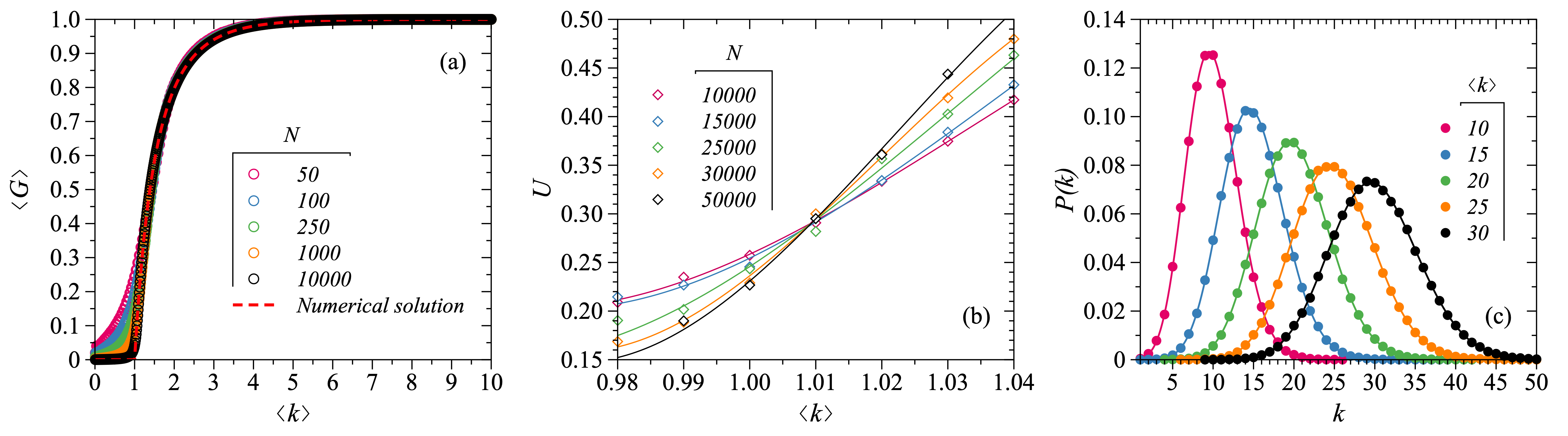}
    \caption{\textbf{Random network connectivity structure.} (a) Average giant component size $\left< G \right>$ versus the average connectivity of the network $\langle k \rangle$ over $10^4$ realizations for different network sizes $N$. The red dashed line indicates the solutions for Eq. \eqref{giantanalytical} for each value of $\langle k \rangle$. (b) Binder fourth-order cumulant of the giant component fraction for several system sizes, intersecting at $\langle k \rangle = \left< k_c \right> = 1.009 \pm 0.003$. The lines are cubic fits to the data points. (c) Degree distributions for several values of $\langle k \rangle$ with $N = 15000$, where the lines are Poisson fits to the data.}
    \label{fig:gc}
\end{figure*}

In this work, we introduce a stochastic consensus model and investigate its dynamics on random networks. We focus on how weak initial asymmetries are amplified into collective agreement and how network connectivity and system size govern the speed and predictability of consensus formation. Through numerical simulations, we characterize the emergence of unanimous states, the decay of the competing state, and the finite-size narrowing of the uncertainty region around the symmetric point. Our findings highlight the potential of the stochastic consensus model as a framework for studying distributed artificial intelligence, autonomous multi-agent systems, blockchain networks, and other decentralized decision-making environments.

This paper is organized as follows. In Sec.~\ref{sec:consensus_net}, we review the fundamental properties of random networks. Section~\ref{sec:model} introduces the stochastic consensus model, informs the simulation protocol, and the quantities that characterize consensus dynamics. Section~\ref{sec:results} presents the numerical results, including the effects of initial conditions, connectivity, and system size on consensus formation. Finally, Sec.~\ref{sec:discussion} summarizes the main findings and outlines future research directions.

\section{\label{sec:consensus_net}Consensus Network}

Effective decentralized decision-making depends fundamentally on the structure of the underlying interaction network through which information, influence, and decisions propagate. Network topology determines how rapidly local information spreads, how efficiently agreement emerges, and how resilient the collective decision process is to delays, failures, or conflicting signals. In general, highly connected networks facilitate information diffusion and accelerate consensus formation, whereas sparse or fragmented networks can slow the propagation of local agreement and hinder the emergence of a coherent collective state \cite{Fadda2022, Musilek2021}.

Given the central role of connectivity in consensus dynamics, we investigate the stochastic consensus model on random networks. Random networks provide a natural baseline for studying decentralized systems because they combine analytical tractability with stochastic interaction patterns while capturing essential aspects of information propagation in large populations of interacting agents \cite{erdos1960, bollobas2001, dall2002, Barabasi2016}. Although real decentralized systems often exhibit more complex topologies, random networks allow the isolation of the effects of connectivity and system size on consensus formation without the additional influence of structural heterogeneity.

A random network is generally specified by two parameters: the number of nodes $N$ and the probability $r$ that any pair of nodes is connected by a link. These parameters determine the average degree $\langle k \rangle = r(N-1)$ and an expected number of links $\langle L \rangle = \langle k \rangle N/2$. In this context, the Erd\"os--R\'enyi model provides a canonical framework for constructing random networks with average connectivity $\langle k \rangle$ by randomly distributing $\langle L \rangle$ connections between nodes while ensuring that no duplicate connections are formed between any pair of nodes \cite{erdos1960}.

Figure \ref{fig:netvis} illustrates the construction of a random network using the Erd\"os--R\'enyi model. Starting with $N = 15$ isolated nodes in Fig. \ref{fig:netvis}(a), Fig. \ref{fig:netvis}(b) illustrates the intermediate step of randomly connecting different nodes by including single links. Figure \ref{fig:netvis}(c) shows the final random network with $\langle k \rangle = 2$, comprising four distinct connected components, with $G_{G} = 1/15$, $G_{O} = 2/15$, $G_{P} = 3/15$ and $G_{B} = 9/15$ representing the fraction of nodes in each group. The blue component is the largest and is referred to as the giant component of the network.

As the average degree $\langle k \rangle$ increases, the network becomes progressively more connected, and the fraction of nodes belonging to the giant component correspondingly grows. In the limit $N \to \infty$, the giant-component fraction $G$ is determined by the self-consistent equation \cite{Barabasi2016}
\begin{equation}
	G = 1-e^{-\langle k \rangle G}.
\label{giantanalytical}
\end{equation}
Random networks undergo a topological phase transition at the critical average degree $\langle k_c \rangle = 1$, where a giant connected component first emerges in the thermodynamic limit. For $\langle k \rangle > 1$, the giant component grows continuously with increasing connectivity and eventually contains most of the network nodes.

In Fig. \ref{fig:gc} (a), we show the average giant component fraction $\langle G \rangle$ versus the average connectivity $\langle k \rangle$ over $10^4$ network realizations for $N$ ranging from $50$ to $10000$. We also include the numerical solution of Eq. \eqref{giantanalytical}, shown as a red dashed line, which is in excellent agreement with the simulation results. In Fig. \ref{fig:gc} (b), we present the fourth-order Binder cumulant of the giant-component fraction $G$, defined as
\begin{equation}
	U = 1-\frac{\langle G^4\rangle}{3\langle G^2\rangle^2},
\end{equation}
where the averages are taken over independent network realizations. The crossing of the curves for different system sizes provides an estimate of the critical connectivity, yielding $\langle k_c \rangle \simeq 1.01$, in agreement with the expected Erd\"os--R\'enyi percolation threshold \cite{Barabasi2016, Binder1981}. 

In Fig. \ref{fig:gc}(c), we display the connectivity distribution for various values of $\langle k \rangle$, using $N = 15000$ nodes. As expected for Erd\"os--R\'enyi networks, the degree distribution follows a binomial form, where most nodes have a similar number of connections located around the peak of the curve, and very few nodes exhibit extremely high or low degrees. For $N \gg \langle k \rangle$, this binomial distribution can be well-approximated by a Poisson distribution, and we fit our results using $P(k) = (\langle k \rangle)^{k}e^{-\langle k \rangle} /{k!}$, which demonstrates quantitative agreement with the data.

\section{stochastic consensus model}
\label{sec:model}

We place an agent on each network node and denote the state of the $i$-th agent at time $t$ by $s_i(t)$. The state represents the agent's current decision and can assume one of two possible values: $+1$ or $-1$. These states may be interpreted as acceptance or rejection of a proposal, support or opposition to an opinion, agreement or disagreement with a decision, or any other binary choice. In this work, we focus exclusively on the collective dynamics of consensus formation that emerge from local interactions among agents, without specifying the underlying origins of individual initial decisions.

The stochastic consensus model (SCM) updates the state of the $i$-th agent by evaluating the states of its neighbors at the instant $t$. We define the local influence on agent $i$ as
\begin{equation}
	\Lambda_{i}(t) = \sum_{\delta \, =\, 1}^{k_{i}} s_{\delta}(t),
\label{sumneighborhood}
\end{equation}
where the sum runs over all $k_i$ neighbors connected to agent $i$. The sign of $\Lambda_i(t)$ captures the predominant state in the local neighborhood: a positive value indicates a majority of neighbors in state $+1$, a negative value indicates a majority in state $-1$, and $\Lambda_i(t) = 0$ corresponds to a tie, with equal numbers of neighbors in states $+1$ and $-1$. We define a local alignment measure as
\begin{equation}
	\gamma_i(t) = -s_{i}(t)\text{sgn}[\Lambda_{i}(t)],
\label{energy}
\end{equation}
where $\textrm{sgn}[x] = -1, 0, +1$ for $x < 0$, $x = 0$ and $x > 0$, respectively. The quantity $\gamma_i(t)$ characterizes the alignment of agent $i$ with the local majority. Thus, the probability that agent $i$ flips its state at time $t$ is defined as
\begin{equation}
	w_{i}(t) = \text{H}[\gamma_i(t)],
\label{wi}
\end{equation}
in which $\textrm{H}(y) = 0, 1/2$ and $1$ for $y < 0$, $y = 0$ and $y >0$, respectively.

The update probability of Eq. \eqref{wi} accounts for the three possible local configurations. When $s_i(t)=\textrm{sgn}[\Lambda_i(t)]$, it yields $\gamma_i(t)=-1$ and $w_i = 0$. Thus, since agent $i$ is aligned with the local majority, it retains its current state. For the case $s_i(t) = -\textrm{sgn}[\Lambda_i(t)]$, $\gamma_i(t)=+1$, producing $w_i = 1$. Here, node $i$ is anti-aligned with the local majority and flips its state. However, when $\Lambda_i(t) = 0$, corresponding to the absence of a local majority, the agent randomizes its state with equal probability. This feature distinguishes the stochastic consensus model from other consensus and opinion-dynamics models in which ties leave the agent unchanged \cite{Fontanari2012} or in which agents update by randomly copying the state of a neighbor \cite{Castellano2009}. By introducing stochastic state revision in locally undecidable configurations, the model explores the state space randomly and avoids the persistence of deadlocked states. The stochastic consensus model can be regarded as a particular case of the majority-vote model with noise, specifically in the zero-noise limit \cite{de1992isotropic}.

In our analysis, only agents belonging to the giant component participate in the consensus dynamics. The expected number of participating nodes in a random network with average connectivity $\langle k \rangle$ is $N_{\mathrm{eff}} = NG(\langle k \rangle)$, where $G(\langle k \rangle)$ denotes the fraction of nodes belonging to the giant component. We exclude agents outside the giant component from the consensus dynamics. For computational convenience, network realizations are regenerated until the number of nodes in the giant component, \(N_{\mathrm{eff}}\), is an integer multiple of 10. For the connectivity regime considered in most simulations, particularly \(\langle k\rangle \geq 10\), the giant component contains nearly all generated nodes, so this constraint produces only a small adjustment to the effective system size.

\subsection*{\label{sec:simulmetric}Simulation Details and Metrics}
\begin{table*}[ht]
 \caption{\label{tab:status}%
 Configuration sets for each case informing the magnetization $m$, configuration $S$, positive fraction $n_+$, and negative fraction $n_-$.}
 \begin{ruledtabular}
 \begin{tabular}{ccccc}
 Status & Magnetization  & Configuration & Positive fraction & Negative fraction\\
 
 \hline\\
 \vspace{1.5mm}
 \textit{Positive unanimity} & $m = +1$ & $S = S_{+\mathbb{1}}$ & $n_+ = 1$ & $n_- = 0$ \\
 \vspace{1.5mm}
 
  \textit{Positive consensus} & $ 0 < m < 1$ & $S = S_{+}$ & $n_+>n_-$ & $n_- < n_+$ \\
  \vspace{1.5mm}
 
 \textit{Polarization} & $ m = 0$ & $S = S_{\oslash}$ & $n_+ = n_-$ & $n_- = n_+$ \\
 \vspace{1.5mm}
 
  \textit{Negative consensus} & $ -1 < m < 0$ & $S = S_{-}$ & $n_+ < n_-$ & $n_- > n_+$ \\
  \vspace{1.5mm}
 
  \textit{Negative unanimity} & $m = -1$ & $S = S_{-\mathbb{1}}$ & $n_+ = 0$ & $n_- = 1$ \\

 \end{tabular}
 \end{ruledtabular}
 \end{table*}

We perform Monte Carlo simulations of the stochastic consensus model on random networks with sizes ranging from $N$, different values of the average connectivity $\langle k \rangle$, and initial concentration of positive nodes $p$. At each elementary update, an agent is selected at random, and its state is updated according to the probability given by Eq. \eqref{wi}. Each such update attempt advances the simulation by one Update Time Step (UTS). We use UTS as the primary time unit because it provides a direct measure of the number of individual node-update attempts required for the system to reach consensus or unanimity. We run simulations up to a total evaluation time $T$.

Each run begins by generating a random network of size, or volume, $N$, and average connectivity $\langle k \rangle$. For the connectivity values considered here, particularly $\langle k\rangle \geq 10$, the giant component essentially contains the entire network. For each network, we randomly distribute a fraction $p$ of agents with state $+1$ at $t_0 = 0$, while the remaining fraction $1 - p$ adopts the state $-1$. This initialization is repeated up to $\mathcal{R}$ times for each network, and up to $\mathcal{N}$ independently generated networks are considered to provide sufficient sampling over both topological and initial-state configurations. 

The subsequent dynamics evolve the initial configuration \begin{equation}
	S(N, \langle k \rangle, p, t_0) = \{s_1(t_0), s_2(t_0), ..., s_N(t_0)\},
\end{equation}
into the time-dependent configuration 
\begin{equation}
	S(N, \langle k \rangle, p, t) = \{s_1(t), s_2(t), ..., s_N(t)\}.
\end{equation}

To characterize how the initial conditions and network properties influence the consensus dynamics, we define an order parameter $m(t)$ inspired by the magnetization per spin in the Ising model. Let $n_+(t)$ and $n_-(t)$ denote the fraction of agents in states $+1$ and $-1$ at time $t$, respectively. Hence,
\begin{equation}
	m(t) = \frac{1}{N_{\mathrm{eff}}} \sum_{i \, =\, 1}^{N_{\mathrm{eff}}} s_i(t) = n_+(t) - n_-(t).
\label{mag}
\end{equation}
We list all possible scenarios for the time-dependent behavior of $m$ and define configuration sets for each case in Table \ref{tab:status}. We denote $S_{\pm \mathbb{1}} = \{\pm 1, \pm 1, ..., \pm 1\}$, where we omit the explicit parameter dependencies for clarity. Note that we distinguish consensus, in which one state constitutes a majority and $m \neq 0$, from unanimity, in which all agents share the same state, yielding $m = \pm 1$.  

We define the unanimity probability, or unanimity rate, $u = u(N, \langle k \rangle, p, t)$, as the fraction of realizations that have reached a unanimous state at time $t$,
\begin{equation}
	u(N, \langle k \rangle, p, t) = \frac{||S_{+\mathbb{1}}|| + ||S_{-\mathbb{1}}||}{||S(N, \langle k \rangle, p, t_0)||},
\label{coneffic}
\end{equation}
where $||\cdot||$ denotes the cardinality operator. This quantity measures the frequency that the system has evolved to either positive unanimity ($m=+1$) or negative unanimity ($m=-1$) by time $t$. To estimate $u$, we average over independent network realizations and initial configurations and count the fraction of runs that reach one of the two unanimous absorbing states.

We also aim to evaluate whether the system's final unanimity reflects its initial configuration or whether it can undergo a reversal. In particular, even if the system is initialized with a majority of agents in state $+1$ (or $-1$), it may ultimately converge to the opposite unanimous state, yielding $m=-1$ (or $m=+1$), indicating a complete flip of the global consensus. To quantify this reversal behavior, we define the unanimity fractions $f_{+\mathbb{1}}$ and $f_{-\mathbb{1}}$, representing the directional probability that the system converges to full $+1$ or $-1$ unanimity, respectively:
\begin{equation}
  f_{\pm {\mathbb{1}}}(N, \langle k \rangle, p, t) = \frac{||S_{\pm \mathbb{1}}||}{||S(N, \langle k \rangle, p, t_0)||}.       
\end{equation}
Naturally, $u = f_{+{\mathbb{1}}} + f_{-{\mathbb{1}}}$. 

We also consider the average values of all the quantities described. For any quantity $q_{\kappa, r}$ evaluated in the network $\kappa$ at the realization of the initial state $r$, its average value $\langle q \rangle$ is estimated by
\begin{equation}
	\langle q \rangle (p, t) = \frac{1}{\mathcal{N} \mathcal{R}} \sum_{\kappa \, =\, 1}^{\mathcal{N}} \sum_{r \, =\, 1}^{\mathcal{R}} q_{\kappa, r}(p, t),
\end{equation}
\label{calcaverages} 

\noindent in which $\mathcal{N}$ stands for the total number of independent networks created and $\mathcal{R}$ indicates the total number of initial configurations for each network realization.

Table~\ref{tab:simParameters} summarizes the system volumes, average connectivity, initial concentration of $+1$ nodes, total simulation time, and averaging measures of the stochastic consensus model on random networks.

\begin{table}[!ht]
\centering
\caption{Simulation parameters and numerical setup}
\label{tab:simParameters}
\begin{ruledtabular}
\begin{tabular}{clc}
Name & Description & Value \\
\hline\\

$N$ & Number of agents & $100$ to $50000$ \\

$ \langle k \rangle$ & Average connectivity per node & $5$ to $1000$ \\

$p$ & Positive fraction initial concentration & $[0.0, 1.0]$ \\

$T$ & Total evolution time & $2 \times 10^6$ \\

$\mathcal{N}$ & Number of networks & up to $300$ \\

$\mathcal{R}$ & Number of runs & up to $500$ \\

\end{tabular}
\end{ruledtabular}
\end{table}

\section{Numerical Results}
\label{sec:results}

\subsection*{Unanimity and initial fraction of positive nodes}
Initially, we explore the effects of the system's initial state by varying the density $p$ of randomly distributed $+1$ agents from $0.50$ to $0.99$. The case $p = 1$ corresponds to an initially unanimous configuration and therefore does not involve consensus formation dynamics. We investigate how many realizations yield a decision by reaching unanimity in a system with $N = 10^4$ agents and average connectivity $\langle k \rangle = 10$ in $2 \times 10^6$ UTS.

Table \ref{tab:tableprobf} presents the mean unanimity times $\langle \tau \rangle$, corresponding to $99\%$ confidence intervals ($CI$), the relative interval width $\Delta$, unanimity fractions $f_{\pm{\mathbb{1}}}$, and fractions of realizations that achieved unanimity $u$ for simulations on $100$ networks and $100$ configurational samples for different values of $p$. To enable an estimation of the relative distance between minimum and maximum unanimity times and provide a direct comparison of results across different realization settings, we define $\Delta$ as
\begin{equation}
\Delta\;=\; \frac{\tau_{\text{upper}}-\tau_{\text{lower}}}{\langle \tau \rangle}, 
\end{equation}
for a confidence interval of $[\tau_{\text{lower}},\,\tau_{\text{upper}}]$. 

\begin{table}[ht]
 \caption{\label{tab:tableprobf}%
 Unanimity time mean values $\langle \tau \rangle$, $99\%$ confidence intervals, $CI$, relative interval width $\Delta$, fraction of realizations that achieved positive and negative unanimity $f_{\pm \mathbb{1}}$, respectively, and unanimity rate $u$ for different $p$ values. The quantities $\Delta$, $f_{\pm\mathbb{1}}$, and $u$ are reported as percentages.}
 \begin{ruledtabular}
 \begin{tabular}{cccccccc}
 $p$ & $\langle \tau \rangle$  & $CI$ & $\Delta$ & $f_{+\mathbb{1}}$ & $f_{-\mathbb{1}}$ & $u$\\
 \hline
 
0.50 & 161539 & [160078, 163000] & 1.8 & 49.76 & 50.22 & 99.98 \\
0.51 & 123020 & [122558, 123482] & 0.8 & 99.42 & 0.58 & 100 \\
0.52 & 110698 & [110357, 111040] & 0.6 & 100 & 0 & 100 \\
0.53 & 105671 & [105339, 106003] & 0.6 & 100 & 0 & 100 \\
0.54 & 102509 & [102175, 102842] & 0.7 & 100 & 0 & 100 \\
0.55 & 100228 & [99892, 100564] & 0.7 & 100 & 0 & 100 \\
0.60 & 94364 & [94025, 94704] & 0.7 & 100 & 0 & 100 \\
0.65 & 90431 & [90102, 90761] & 0.7 & 100 & 0 & 100 \\
0.70 & 87795 & [87461, 88130] & 0.8 & 100 & 0 & 100 \\
0.75 & 85213 & [84880, 85546] & 0.8 & 100 & 0 & 100 \\
0.80 & 82574 & [82242, 82905] & 0.8 & 100 & 0 & 100 \\
0.85 & 79285 & [78958, 79613] & 0.8 & 100 & 0 & 100 \\
0.90 & 75341 & [75006, 75677] & 0.9 & 100 & 0 & 100 \\
0.95 & 68146 & [67812, 68479] & 1.0 & 100 & 0 & 100 \\
0.99 & 51999 & [51671, 52327] & 1.3 & 100 & 0 & 100 \\

 \end{tabular}
 \end{ruledtabular}
 \end{table}
 
At the symmetric initial distribution $p = 0.50$, the unanimity rate $u$ is equal to $99.98\%$, with $f_{+ {\mathbb{1}}}$ and $f_{-{\mathbb{1}}}$ each occurring with approximately $50\%$ probability. This setup also produced the longest average unanimity time, $\langle \tau \rangle = 161539$ UTS. As $p$ increases slightly above $0.50$, a sharp transition is observed, and at $p = 0.51$, the fraction $f_{+ \mathbb{1}}$ of positive-unanimity outcomes in $2 \times 10^6$ UTS rises abruptly to $99.42\%$, indicating a robust final decision aligned with the initially dominant state. Across all investigated values of $p$, the small values of $\Delta$ show that the model reaches unanimity within a consistently narrow temporal interval.

This behavior highlights a remarkable resilience of the stochastic consensus dynamics in which even an exceedingly small bias in the initial distribution is amplified into an overwhelmingly consistent final decision. The system effectively suppresses fluctuations around the symmetric point, locking onto the majority direction with near-deterministic reliability. Consistent with this trend, all realizations reach positive unanimity for $p \geq 0.52$ in the present network configuration.

\begin{figure}[H]
    \centering
    \hspace*{-1cm}
    \includegraphics[width=1\linewidth]{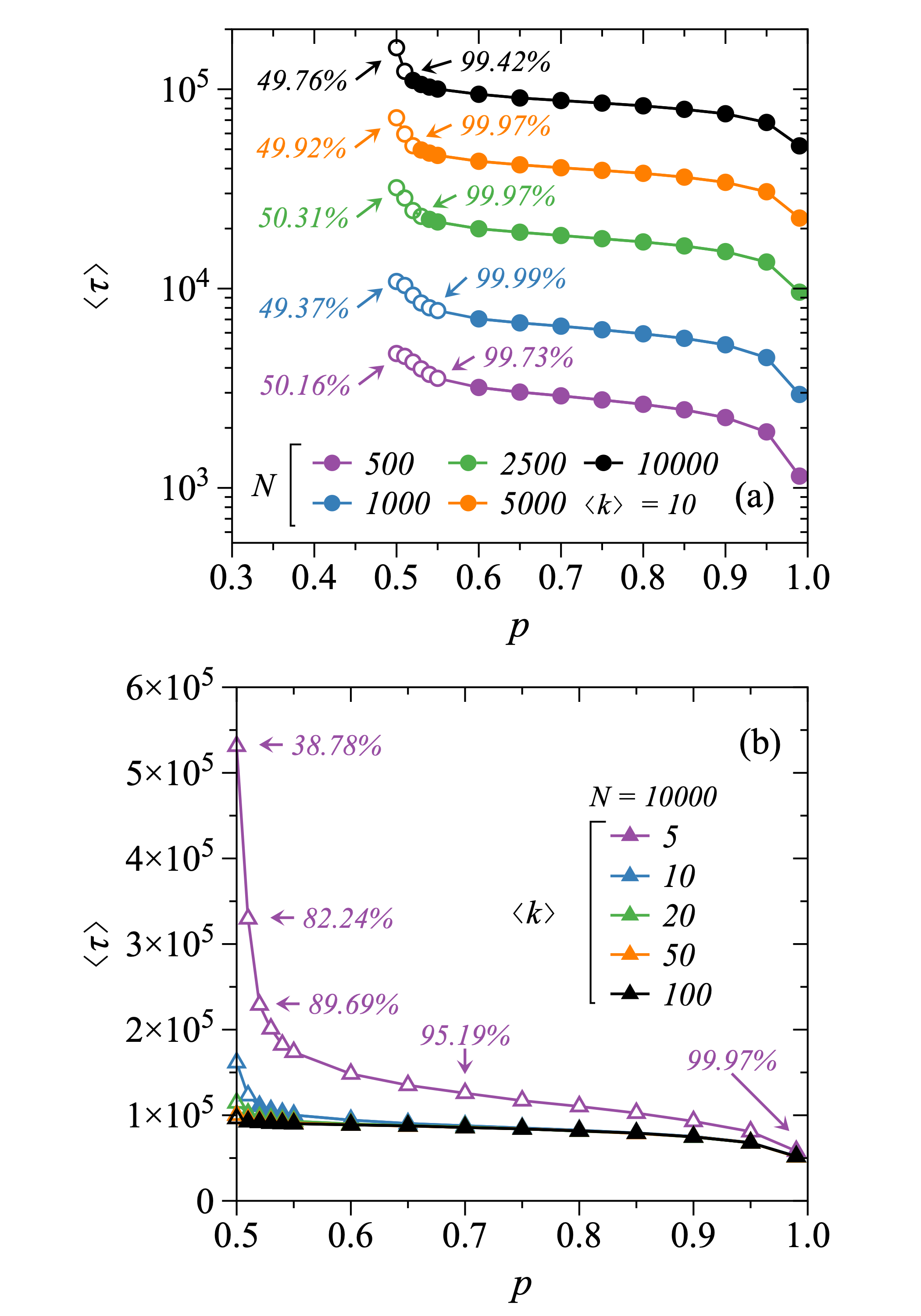}
    \caption{
    \textbf{Unanimity convergence for different initial conditions.} Average unanimity time versus the initial fraction of $+1$ agents for (a) several system volumes $N$ and $\langle k \rangle = 10$ and (b) different $\langle k \rangle$ values and $N = 10^4$. Closed symbols represent realizations in which $100\%$ of the samples reached positive unanimity in $2 \times 10^6$ UTS. Lines are guides to the eye.}
    \label{fig:ctime_vs_p_varNk}
\end{figure}

These results suggest that the proposed consensus scheme can be viewed as a \textit{trust amplifier} for decentralized decision-making with deadlock-breaking capability. If the system contains up to $49\%$ of agents that may behave adversarially or follow a corrupted signal, the dynamics still drive the network toward a unanimous decision aligned with the small initial uncorrupted majority ($p = 0.51$) in more than $99\%$ of cases.

Figure \ref{fig:ctime_vs_p_varNk} illustrates how the average unanimity time $\langle \tau \rangle$ depends on the initial fraction $p$ of agents in state $+1$ for networks of different volumes $N$ and average connectivities $\langle k \rangle$ in $2 \times 10^6$ UTS. In Fig. \ref{fig:ctime_vs_p_varNk}(a) the plot includes curves for $\langle k \rangle = 10$ and $N = 500, 1000, 2500, 5000$, and $10000$, each represented by a different color. The vertical axis spans roughly three orders of magnitude in $\langle \tau \rangle$. Open symbols denote values of $p$ for which some realizations did not achieve positive unanimity, and the percentages annotated near these points indicate the fraction of networks that did converge to the $+1$ state.

Under symmetric initial conditions, that is, $p = 0.50$, the system exhibits the longest unanimity times, reflecting the critical nature of balanced initial states. In addition, larger system sizes require longer times to reach positive unanimity and exhibit a similar qualitative dependence on $p$. This result suggests that increasing the system size does not introduce abrupt qualitative changes in the model's behavior, but rather preserves the underlying mechanisms governing consensus formation. We also observe that the mean unanimity time decreases monotonically with $p$, indicating that stronger initial agreement leads to faster convergence. The tight relative intervals $\Delta$ across all values of $p$ indicate low statistical dispersion in the estimated mean unanimity times.

Figure \ref{fig:ctime_vs_p_varNk}(b) displays the dependence of the average unanimity time, $\langle \tau \rangle$, on the initial fraction $p$ of agents in state $+1$ for networks with $N = 10^4$ and average connectivities $\langle k\rangle = 5, 10, 20, 50$, and $100$. For networks with low average degree, $\langle k \rangle = 5$, the system exhibits dramatically longer unanimity times near symmetric initial conditions $p \approx 0.50$, with $\langle \tau \rangle$ exceeding $5 \times 10^5$ UTS. This low-connectivity scenario also shows a slightly lower positive-unanimity rate, with about $99.97\%$ of realizations reaching positive unanimity at $p = 0.99$. For $\langle k \rangle \geq 10$, the curves flatten considerably, indicating that these networks display a much smaller slowdown and achieve near-uniform consensus probabilities. As $p$ grows beyond $0.51$, all curves exhibit similar unanimity times regardless of $\langle k \rangle$. This behavior indicates that, beyond this initial asymmetry, increasing connectivity above $\langle k \rangle = 10$ does not substantially accelerate convergence in networks with $10^4$ agents.
  

\subsection*{Average connectivity effects}
The number of connections critically shapes the evolution of collective phenomena that emerge in networked models. Higher average connectivity creates a more interconnected interaction structure, facilitating the propagation of local information and accelerating the formation of agreement. In this context, the average connectivity $\langle k \rangle$ directly influences the speed of consensus formation, with more highly connected networks generally reaching unanimity more rapidly.

\begin{table}[ht]
 \caption{\label{tab:tableprobk}%
 Unanimity time mean values, confidence intervals, and the fraction of realizations that achieved consensus for $100$ networks and $100$ configurational samples with $N = 10^4$ and $p = 0.51$ for several values of the average connectivity $\langle k \rangle$ over $2\times 10^6$ UTS. The quantities $\Delta$, $f_{\pm\mathbb{1}}$, and $u$ are reported as percentages.}
 \begin{ruledtabular}
 \begin{tabular}{cccccccc}
 $\langle k \rangle$ & $\langle \tau \rangle$  & $CI$ & $\Delta$ & $f_{+\mathbb{1}}$ & $f_{-\mathbb{1}}$ & $u$\\
 \hline
5  & 333291 & [328710, 337873] & 2.7 & 85.63 & 2.05 & 87.68 \\
6  & 214444 & [211959, 216929] & 2.3 & 95.75 & 1.97 & 97.72 \\
7  & 167498 & [165809, 169187] & 2.0 & 97.91 & 1.48 & 99.39 \\
8  & 144285 & [143281, 145288] & 1.4 & 98.71 & 1.27 & 99.98 \\
9  & 131397 & [130788, 132005] & 0.9 & 99.23 & 0.76 & 99.99 \\
10 & 123020 & [122558, 123482] & 0.8 & 99.42 & 0.58 & 100\\
20 & 101613 & [101275, 101951] & 0.7 & 99.95 & 0.05 & 100 \\
30 & 97502 & [97166, 97838] & 0.7 & 99.98 & 0.02 & 100 \\
40 & 95749 & [95415, 96083] & 0.7 & 100 & 0 & 100 \\
50 & 94662 & [94333 94991] & 0.7 & 100 & 0 & 100\\
100	& 92999 & [92666, 93332] & 0.7 & 100 & 0 & 100\\
500	& 91065 & [90737, 91394] & 0.7 & 100 & 0 & 100\\
1000 &	90643 &	[90316, 90969] & 0.7 & 100 & 0 & 100\\

 \end{tabular}
 \end{ruledtabular}
 \end{table}
 
\begin{figure}[ht]
    \centering
    \hspace*{-1cm}
    \includegraphics[width=1\linewidth]{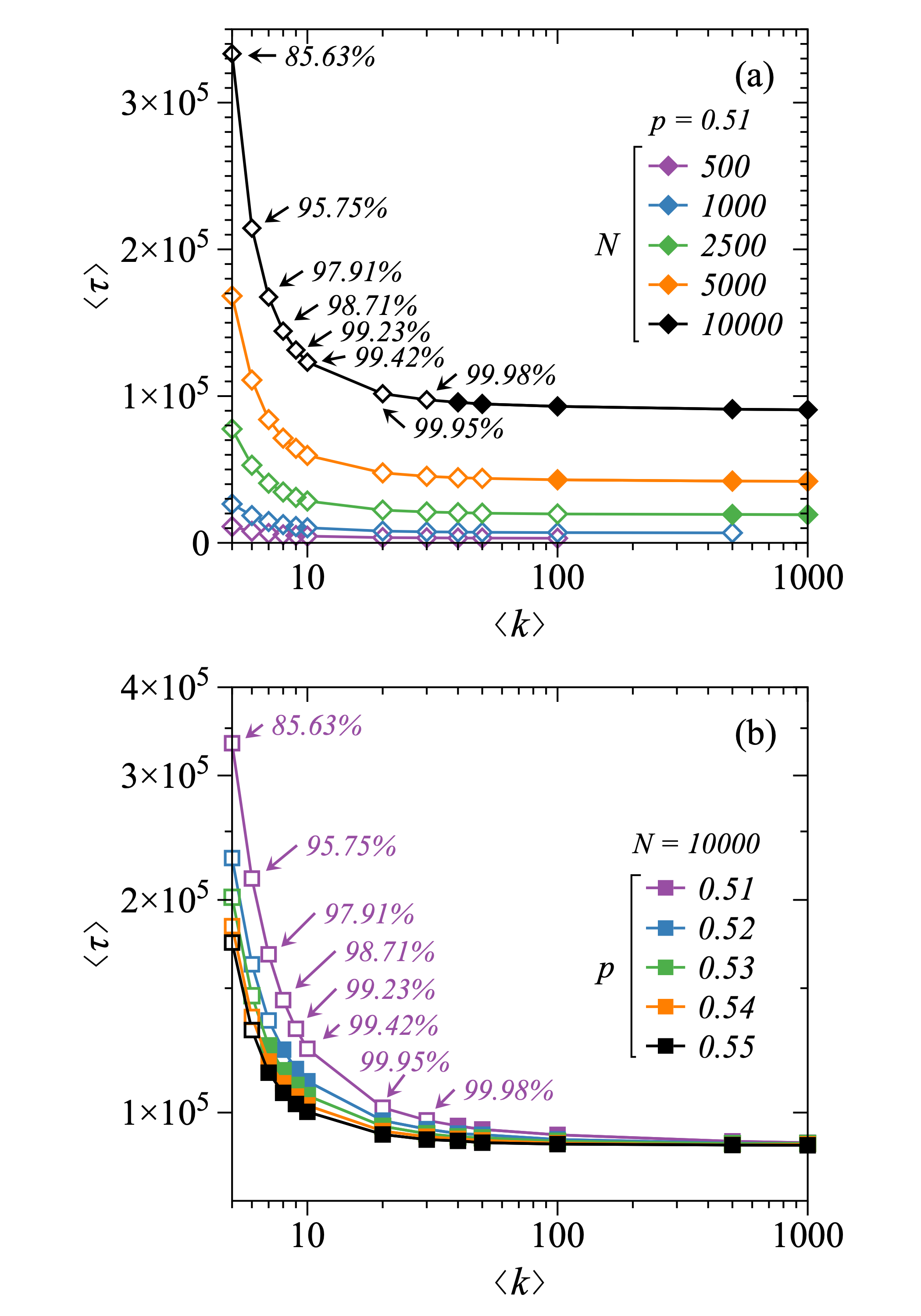}
    \caption{
    \textbf{Time for unanimity convergence for different average connectivities.} Unanimity time versus average connectivity $\langle k \rangle$ for (a) several system volumes $N$ with $p = 0.51$, and (b) different values of $p$ with $N = 10^4$ in $2 \times 10^6$ UTS. Closed symbols represent realizations in which $100\%$ of the samples reached positive unanimity. Lines are guides to the eye.}
    \label{fig:ctime_vsk_varNp}
\end{figure}

Table \ref{tab:tableprobk} reveals a strong influence of network connectivity on the consensus dynamics. As the average connectivity increases from $\langle k \rangle = 5$ to $\langle k \rangle = 1000$, the mean unanimity time $\langle \tau \rangle$ decreases by nearly a factor of four, indicating a substantial acceleration of consensus formation. Despite the small initial bias ($p=0.51$), the frequency of positive unanimity steadily increases from $85.63\%$ to $100\%$ with the increase of $\langle k \rangle$, demonstrating that highly connected networks reach consensus both faster, while the occurrence of negative unanimity vanishes. For $\langle k \rangle \geq 40$, all sampled realizations converge to positive unanimity, highlighting the amplification of weak initial majorities into robust collective decisions.

Figure \ref{fig:ctime_vsk_varNp}(a) displays $\langle \tau \rangle$ as a function of $\langle k \rangle$ for fixed initial positive density $p = 0.51$ and different system sizes $N = 500, 1000, 2500, 5000$, and $10000$ in $2 \times 10^6$ UTS. For all system sizes, the unanimity time decreases rapidly as $\langle k \rangle$ increases, indicating that higher connectivity strongly accelerates the convergence toward consensus. This decrease in unanimity time is particularly pronounced for sparse networks ($\langle k \rangle \lesssim 10$), where increasing connectivity significantly enhances collective agreement. For larger connectivities, $\langle \tau \rangle$ approaches a plateau, suggesting a saturation regime in which further increases in $\langle k \rangle$ yield only marginal dynamical speed-up. The percentages indicated by arrows correspond to the fraction of realizations that reached positive unanimity within the simulation window. 

Figure \ref{fig:ctime_vsk_varNp}(b) shows $\langle \tau \rangle$ as a function of $\langle k \rangle$ for a fixed system size $N = 10^4$ and different initial densities $p = 0.51$ to $0.55$. Increasing the initial bias toward one opinion systematically reduces the time to positive unanimity, especially in the low-connectivity regime. However, for large $\langle k \rangle$, all curves collapse onto a common plateau, indicating that network connectivity dominates over initial conditions in determining the consensus timescale in dense networks. These results confirm that higher connectivity enhances both the robustness and efficiency of unanimity formation in decentralized consensus systems, while sparse networks can suffer from fragmentation and slow convergence. In practice, this implies that designing network topologies with moderate to high average connectivity is essential to achieving fast, reliable consensus in such decentralized environments.

\subsection*{Consensus evolution and volume dependency}

\begin{table}[ht]
 \caption{\label{tab:tableprobN}%
 Unanimity time mean values, confidence intervals, and the fraction of realizations that achieved consensus for $100$ networks and $100$ configurational samples with varying system size $N$, $\langle k \rangle = 10$ and $p = 0.51$ in $2 \times 10^6$ UTS. The quantities $\Delta$, $f_{\pm\mathbb{1}}$, and $u$ are reported as percentages.}
 \begin{ruledtabular}
 \begin{tabular}{cccccccc}
 $N$ & $\langle \tau \rangle$  & $CI$ & $\Delta$ & $f_{+\mathbb{1}}$ & $f_{-\mathbb{1}}$ & $u$\\
 \hline
100  & 627 & [621, 633] & 1.9 & 60.31 &	39.56 & 99.87 \\
200  & 1525	& [1508, 1542] & 2.2 & 63.16 & 36.77 & 99.93 \\
500   & 4566 & [4525, 4608] & 1.8 & 70.98 & 28.97 & 99.95 \\
1000  & 10299 & [10212, 10386] & 1.7 & 78.14 & 21.84 & 99.98 \\
2500  & 28487 & [28288, 28686] & 1.4 & 89.49 & 10.51 & 100 \\
5000  & 59837 & [59410, 60265] & 1.4 & 95.88 & 4.12  & 100 \\
10000 & 123020 & [122558, 123482] & 0.8 & 99.42 & 0.58 & 100 \\
20000 & 257024 & [256297, 257751] & 0.6 & 99.98 & 0.02  & 100 \\
30000 & 396080 & [395047, 397112] & 0.5 & 100   & 0		& 100 \\
40000 &	539774 & [538389, 541158] & 0.5 & 100 & 0 & 100\\
50000 &	684501 & [682788, 686214] &	0.5 & 100 & 0 & 100\\
  
 \end{tabular}
 \end{ruledtabular}
 \end{table}
 
We perform a systematic exploration of the unanimity statistics under varying system sizes, and Table \ref{tab:tableprobN} summarizes the results for systems with fixed average connectivity $\langle k \rangle = 10$ and initial density $p = 0.51$. For each system size, results are averaged over $100$ independent network realizations and $100$ configurational samples, with a maximum observation window of $2 \times 10^6$ update time steps. Overall, our results indicate that larger systems take longer to converge: the average unanimity time increases monotonically with system size, reflecting the greater communicative load required for global agreement in larger systems with a slightly greater initial positive state. This growth is consistent with expectations from information propagation in networked systems.

\begin{figure}[ht]
    \centering
    \hspace*{-1cm}
    \includegraphics[width=1\linewidth]{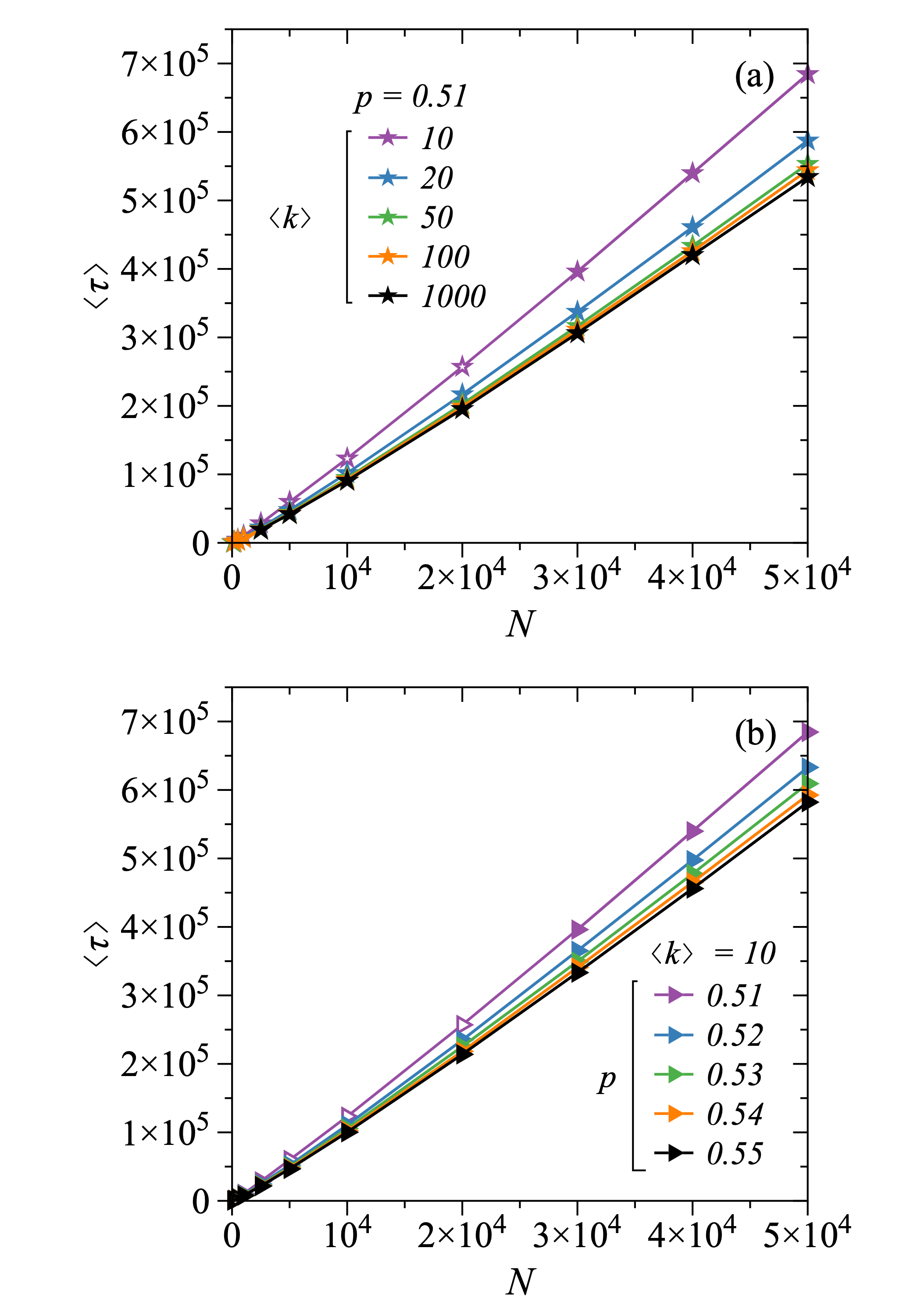}
    \caption{
    \textbf{Unanimity convergence for different system volumes.} Unanimity time versus system volume $N$ for (a) several average connectivities $\langle k \rangle$ with $p = 0.51$ and (b) different values of $p$ with $\langle k \rangle = 10$. Closed symbols represent realizations in which $100\%$ of the samples reached positive unanimity in $2 \times 10^6$ UTS. Lines are guides to the eye.}
    \label{fig:ctime_vsN_varkp}
\end{figure}

For small system sizes, a noticeable fraction of realizations end in either absorbing state, reflecting stronger finite-size fluctuations. As $N$ increases, the final consensus state becomes dominated by the $+1$ opinion, consistent with the slight initial bias $p = 0.51$. Correspondingly, the total unanimity fraction $u$ approaches unity, indicating that for sufficiently large systems all realizations reach consensus within the simulation window. The positive unanimity fraction $f_{+\mathbb{1}}$ also displays size-dependent behavior. While small systems, such as $N = 100$, reach positive unanimity in $60.31\%$ of realizations, $f_{+\mathbb{1}}$ increases steadily with $N$, reaching $f_{+\mathbb{1}} \approx 100\%$ for $N \geq 10000$.

Figure \ref{fig:ctime_vsN_varkp}(a) shows the dependence of the mean unanimity time $\langle \tau \rangle$ on the system size $N$ for several values of the average connectivity $\langle k \rangle$ at fixed $p=0.51$. In all cases, $\langle \tau \rangle$ increases approximately linearly with $N$, indicating that the computational effort required to reach consensus scales proportionally to the number of participating agents. Increasing the average connectivity substantially reduces the unanimity time, particularly when moving from $\langle k \rangle = 10$ to $\langle k \rangle = 20$. For larger connectivities, however, the curves progressively collapse, suggesting diminishing returns in consensus acceleration once the network becomes sufficiently connected.

Figure \ref{fig:ctime_vsN_varkp}(b) presents the corresponding results for different values of the initial fraction $p$ at fixed connectivity $\langle k \rangle = 10$. The unanimity time remains approximately linear in $N$ for all values of $p$, confirming the robustness of the scaling behavior. Increasing the initial bias decreases the time required to reach consensus, as a larger initial majority reduces the number of state changes necessary for unanimity. The magnitude of this effect is comparable to that obtained by increasing the average connectivity, suggesting that both the network topology and the initial state distribution dominate the speed of consensus formation.

\subsection*{Ensemble-averaged consensus dynamics}
\begin{figure*}[ht]
    \centering
    \includegraphics[width=1.0\linewidth]{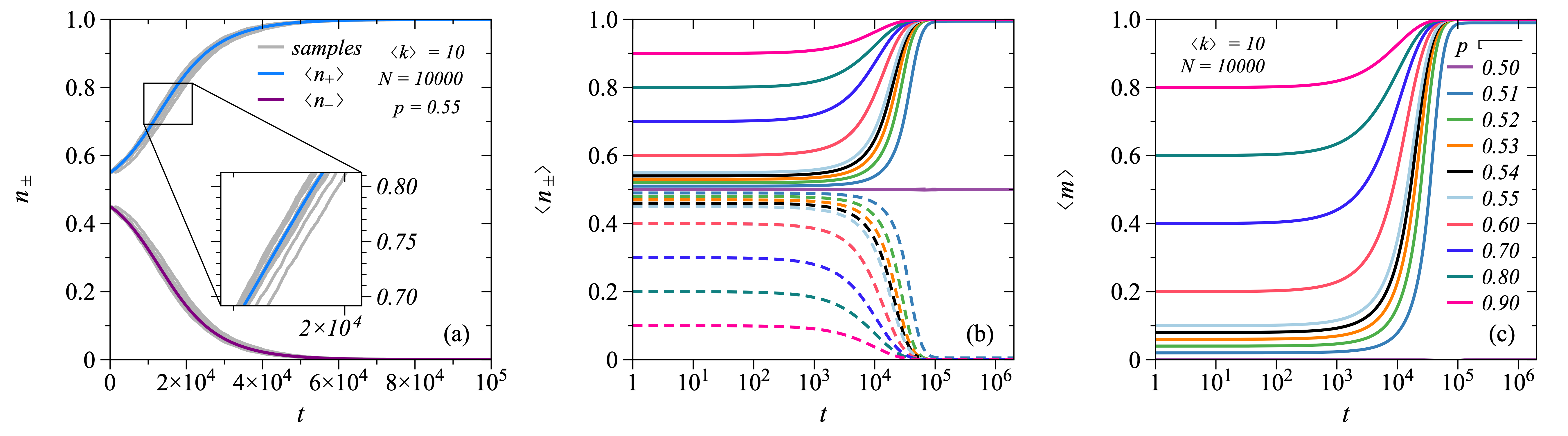}
    \caption{\textbf{Consensus formation and suppression of dissent.} Averaged time evolution of the stochastic consensus dynamics on random networks with $N=10^4$ and average degree $\langle k\rangle=10$. (a) Individual realizations (gray curves) together with the ensemble averages of the fractions of nodes in the $+1$ state, $\langle n_{+}\rangle$ (blue), and the $-1$ state, $\langle n_{-}\rangle$ (purple) for $p=0.55$. (b) Ensemble averages of $\langle n_{+}\rangle$ (solid lines) and $\langle n_{-}\rangle$ (dashed lines) for different values of the initial concentration $p$. (c) Corresponding evolution of the average consensus level $\langle m\rangle$. Increasing $p$ strengthens the initially dominant state, accelerates the decay of the competing state, and drives the system more rapidly toward unanimous consensus.}
    \label{fig:samples_and_averages}
\end{figure*}

To complement the analysis of unanimity times and consensus probabilities, we investigate ensemble-averaged quantities that characterize the collective evolution of the stochastic consensus dynamics. In particular, we consider the average fractions of agents in states $+1$ and $-1$, denoted by $\langle n_{+}\rangle$ and $\langle n_{-}\rangle$, respectively, and the average magnetization $\langle m\rangle$. These quantities provide a macroscopic description of how local interactions progressively amplify initial asymmetries, suppress dissenting opinions, and drive the system toward unanimous consensus.

Figure~\ref{fig:samples_and_averages}(a) compares individual realizations with the corresponding ensemble averages for $N = 10^4$, $\langle k\rangle = 10$, and $p = 0.55$. Although individual trajectories exhibit stochastic fluctuations, the averaged quantities evolve smoothly and reveal a systematic increase in the majority fraction accompanied by a continuous decline of the minority fraction. Fig.~\ref{fig:samples_and_averages}(b) reveals the temporal evolution of $\langle n_{+}\rangle$ and $\langle n_{-}\rangle$ for different values of $p$. As the initial positive bias increases, the majority state gains support more rapidly, while the minority population decreases. The average magnetization of Fig.~\ref{fig:samples_and_averages}(c) also shows this behavior. Higher values of $p$ correspond to larger initial magnetizations and smoother transitions toward $\langle m\rangle = 1$, demonstrating that even modest asymmetries in the initial state are efficiently amplified by the dynamics.

A special case occurs at the symmetry point $p = 0.50$. In this case, neither opinion is initially favored, and the dynamics are invariant under the exchange of the two states. As a consequence, the ensemble averages satisfy $\langle n_+\rangle \approx \langle n_-\rangle \approx 1/2$ and $\langle m\rangle \approx 0$ at all times. As seen in the previous analysis, this behavior should not be interpreted as evidence against consensus formation. Individual realizations still evolve toward either positive or negative unanimity, but both outcomes occur with approximately equal probability. The apparent stationary state observed in the averages results from the restoration of symmetry across the ensemble rather than from the persistence of disorder within individual realizations.

\begin{figure*}[ht]
    \centering
\includegraphics[width=1.0\linewidth]{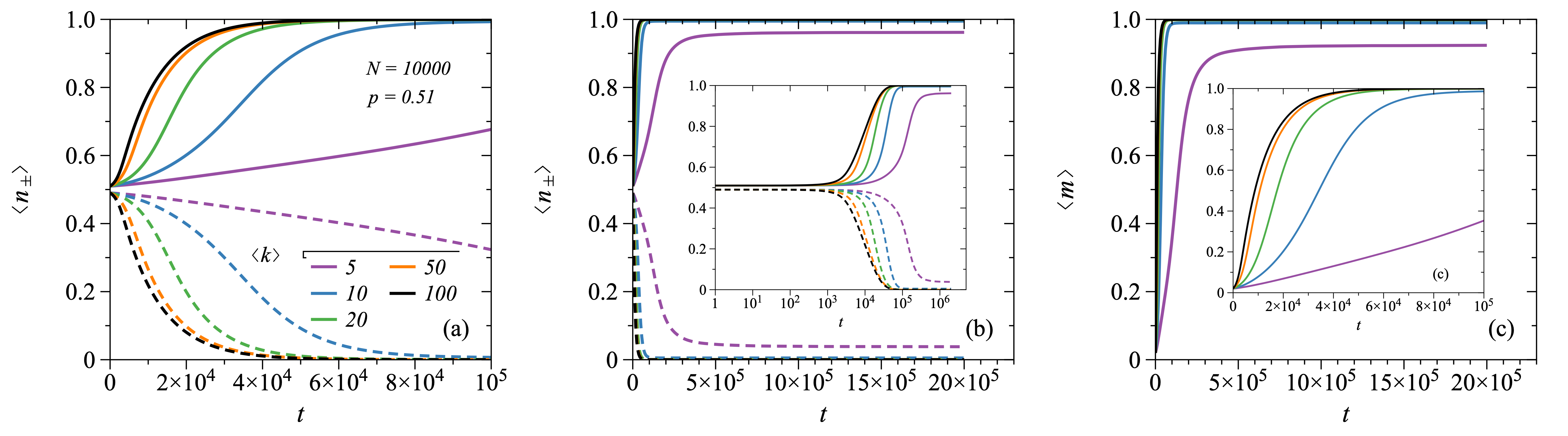}
    \caption{\textbf{Effect of network connectivity on consensus formation.} Averaged time evolution of the stochastic consensus dynamics on random networks with $N=10^4$ and initial concentration $p=0.51$. (a) and (b) Ensemble averages of the fractions of nodes in the $+1$ state, $\langle n_{+}\rangle$ (solid lines), and the $-1$ state, $\langle n_{-}\rangle$ (dashed lines), for different average connectivities $\langle k\rangle$. (c) Corresponding evolution of the average consensus level $\langle m\rangle$. Increasing the average connectivity accelerates the propagation of local agreement throughout the network, promoting a faster growth of the dominant state and a more rapid convergence toward unanimous consensus. Insets show different scales of the same quantity.}
    \label{fig:nmvstvarkp051}
\end{figure*}

In Figure \ref{fig:nmvstvarkp051}, we illustrate the influence of the network connectivity on the averaged consensus dynamics for a fixed bias parameter $p = 0.51$ and system volume $N = 10^4$. Fig. \ref{fig:nmvstvarkp051}(a) presents the time evolution of the average fractions of agents supporting the majority opinion, $\langle n_{+}\rangle$ (solid lines), and the minority opinion, $\langle n_{-}\rangle$ (dashed lines), for different average degrees $\langle k\rangle$. For a fixed $p$, the convergence rate depends strongly on the network connectivity. Increasing $\langle k\rangle$ accelerates the elimination of dissent and promotes faster growth of the majority fraction, reflecting the enhanced information flow and influence propagation in more densely connected networks.

Figure \ref{fig:nmvstvarkp051}(b) shows the results for panel (a) in an extended temporal window of $2 \times 10^6$ UTS, revealing the long-time evolution of the consensus process for the different average connectivities. In the inset, we show the linear-log plot of the same quantity. For $\langle k\rangle \geq 10$, the system rapidly reaches positive unanimity, with $\langle n_{+}\rangle\rightarrow 1$ and $\langle n_{-}\rangle\rightarrow 0$, indicating that almost all realizations converge to the same final state. In contrast, for the sparse network with $\langle k\rangle = 5$, consensus formation remains substantially slower. Furthermore, even after $2 \times 10^6$ UTS, the ensemble averages saturate at $\langle n_{+}\rangle \approx 0.96$ and $\langle n_{-}\rangle \approx 0.04$, demonstrating that a finite fraction of realizations has not yet reached positive unanimity. This slower consensus rate and persistent deviation from the absorbing state reflect the reduced efficiency of information propagation in weakly connected networks.

\begin{figure*}[ht]
    \centering
\includegraphics[width=1.0\linewidth]{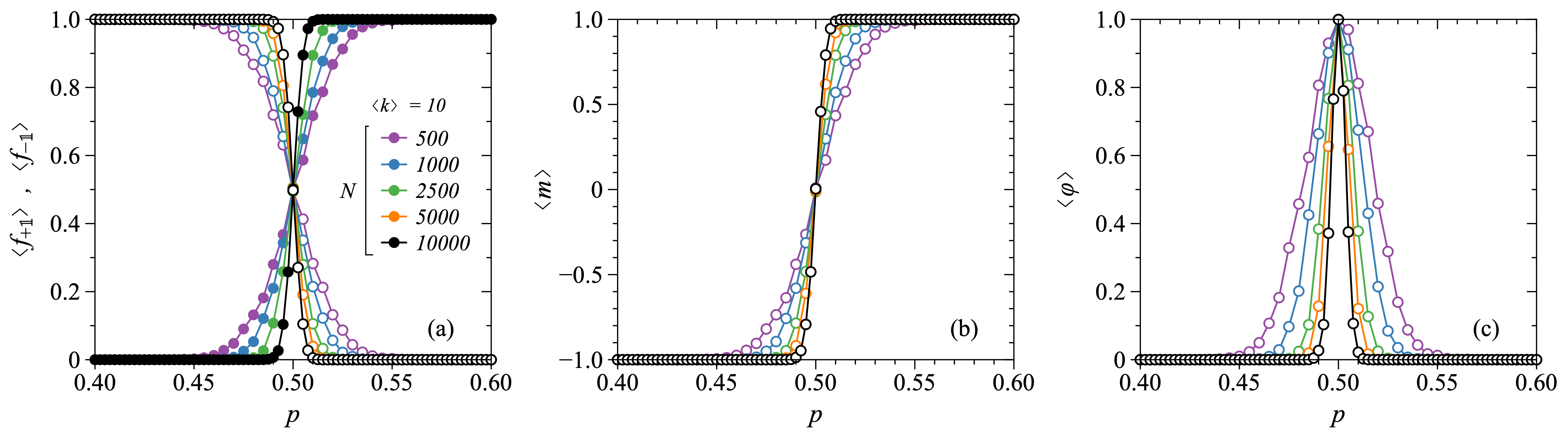}
    \caption{\textbf{Dependence of consensus outcomes on the initial positive concentration.} Consensus evolution versus the initial positive concentration $p$ for systems with average degree $\langle k\rangle = 10$ and $N = 500, 1000, 2500, 5000$ and $10000$. (a) Average unanimity fractions of the $+1$ and $-1$ states, $f_{+\mathbb{1}}$ (filled symbols) and $f_{-\mathbb{1}}$ (open symbols). (b) Average magnetization $\langle m \rangle$. (c) Average consensus uncertainty $\langle \varphi \rangle$. All quantities exhibit a transition at $p_c = 0.50$. Increasing the system size sharpens the transition region, while the peak of $\langle \varphi \rangle$ becomes progressively narrower, indicating convergence toward a singular transition in the thermodynamic limit $N \to \infty$.}
    \label{fig:allversusp}
\end{figure*}

In Fig. \ref{fig:nmvstvarkp051}(c), we show the corresponding evolution of the average magnetization $\langle m\rangle$. Results are similar to those for $\langle n_{+}\rangle$ and $\langle n_{-}\rangle$ in panel (a), as expected, and the ordering process becomes progressively faster as $\langle k\rangle$ increases. These results also confirm that connectivity primarily controls the timescale of consensus formation rather than its ultimate outcome, with highly connected networks reaching full or partial agreement substantially earlier than sparse ones.

\subsection*{Finite-size effects and decision uncertainty}

To further investigate how the stochastic consensus model amplifies small initial asymmetries near the critical region of $p \approx 0.50$, we examine the dependence of the averaged unanimity fractions, and quantify the decision uncertainty associated with the final unanimity state as
\begin{equation}
	\varphi (N, \langle k \rangle) = 4f_{\mathbb{+1}}f_{\mathbb{-1}}.
\end{equation}
The quantity $\varphi (N, \langle k \rangle)$ measures the degree of competition between the two possible unanimity outcomes. It vanishes when the dynamics selects a single unanimity state, for instance, $f_{\mathbb{+1}} = 1$, $f_{\mathbb{-1}} = 0$, and reaches a maximum, $\varphi = 1$, when $f_{\mathbb{+1}} = f_{\mathbb{-1}} = 0.50$, corresponding to maximal uncertainty, where both unanimity states are equally likely. Thus, $\varphi$ measures unanimity ambiguity, analogous to the variance of a binary random variable, and can be interpreted as an indicator of how strongly the system is polarized between the two competing absorbing states.

Figure \ref{fig:allversusp} presents the average fractions $\langle f_{+\mathbb{1}}\rangle$ and $\langle f_{-\mathbb{1}}\rangle$, magnetization $\langle m\rangle$, and decision uncertainty $\langle \varphi \rangle$ for $\langle k \rangle = 10$ and different system volumes $N$ in $2 \times 10^6$ UTS. In Fig. \ref{fig:allversusp}(a), we plot the average fractions of realizations $\langle f_{\pm \mathbb{1}}\rangle$ that reach the $+1$ and $-1$ unanimity states. For $p < 0.50$, the system predominantly reaches the absorbing state $-1$, whereas for $p > 0.50$, the $+1$ unanimity state becomes dominant. At the critical (symmetry) point $p = p_{c} = 0.50$, both absorbing states are equally likely, yielding $\langle f_{+\mathbb{1}}\rangle \approx \langle f_{- \mathbb{1}}\rangle \approx 0.50$. As $N$ increases, the crossover between the two consensus outcomes becomes sharper, suggesting that $p_{c} = 0.50$ separates two deterministic consensus regimes.

Fig. \ref{fig:allversusp}(b) shows the corresponding average magnetization, which changes continuously from $m = -1$ to $m = +1$ as $p$ crosses $p_{c} = 0.50$. For finite systems, this change occurs over a finite interval of $p$, but the transition region narrows with increasing $N$. In the limit $N \to \infty$, the curves approach a step-like behavior, suggesting that an infinitesimal bias is sufficient to determine the final averaged consensus state. Fig. \ref{fig:allversusp}(c) displays the average decision uncertainty $\langle \varphi \rangle$, which is maximal at $p = 0.50$, where both outcomes are equally probable, and vanishes when one consensus state is selected with probability one.

The sharpening of the $\langle \varphi \rangle$ curves with increasing system size indicates that the range of initial conditions yielding uncertain outcomes progressively narrows as the network grows. Since the peak height remains approximately constant and close to unity for all investigated system volumes, the dominant finite-size effect is the narrowing of the consensus uncertainty region surrounding the symmetry point $p_c = 0.50$. To quantify this effect, we define the uncertainty width $W_{\varphi}$ as the full width at half maximum of $\langle \varphi \rangle$,
\begin{equation}
W_{\varphi} = p_{R}-p_{L},
\end{equation}
where $p_L$ and $p_R$ are the values of $p$ satisfying
\begin{equation}
\langle \varphi(p_L)\rangle = \langle \varphi(p_R)\rangle = \frac{1}{2}\langle \varphi\rangle_{\max}.
\end{equation}

The uncertainty width provides a direct estimate of potential final states. Large values of $W_{\varphi}$ indicate a broad interval of initial conditions for which either positive or negative unanimity may emerge, whereas small values correspond to highly predictable unanimity formation. The volume dependence allows us to propose the scaling relation
\begin{equation}
W_{\varphi} \sim N^{-\rho},
\end{equation}
where $\rho = \rho(\langle k \rangle)$ is a volumetric reliability exponent that quantifies the rate at which consensus uncertainty is suppressed as the network volume increases. This relation has a direct practical interpretation: if the network size increases by a factor $\mu$, the uncertainty region decreases by a factor $\mu^{-\rho}$. Consequently, larger decentralized systems require progressively smaller initial asymmetries to reliably determine the final unanimity outcome.

The exponent $\rho$ can be estimated from the slope of a linear fit of $\log W_{\varphi}$ versus $\log N$, providing a quantitative measure of the majority amplification capability of the stochastic dynamics. That is
\begin{equation}
\log W_{\varphi} \sim -\rho \log N.
\end{equation}

\begin{figure}[ht]
    \centering
    \hspace*{-1cm}
    \includegraphics[width=0.85\linewidth]{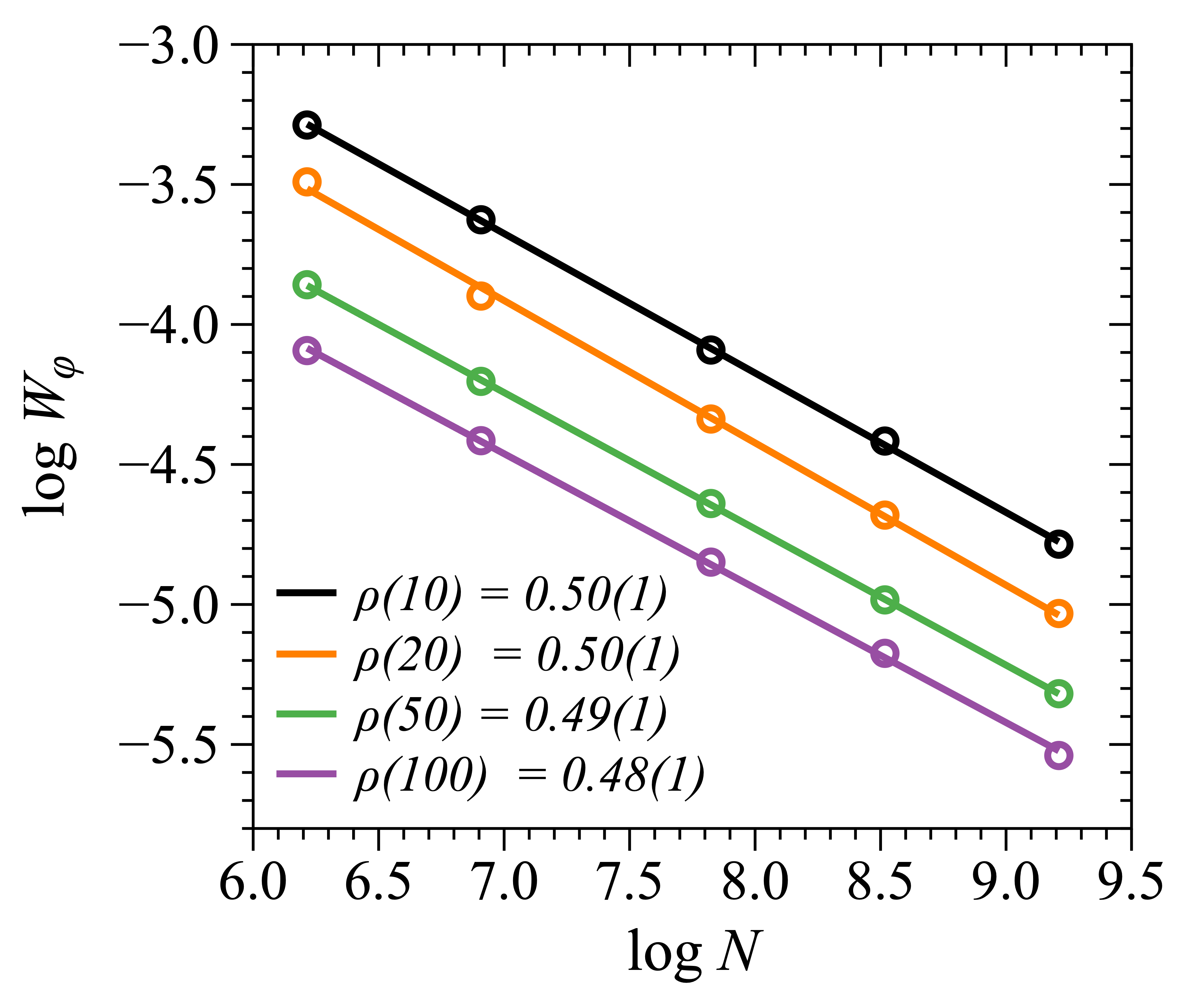}
    \caption{
    \textbf{Finite-size scaling of the uncertainty width.} The linear behavior in the log-log representation indicates a power-law dependence on system volume, $W_{\varphi} \sim N^{-\rho}$. The solid line corresponds to a linear regression, yielding $\rho \approx 0.50$ for all $\langle k \rangle$ investigated.}
    \label{fig:logpN}
\end{figure}

In Fig. \ref{fig:logpN}, we present the dependence of the uncertainty width $W_{\varphi}$ on the system volume $N$ for $\langle k \rangle = 10, 20, 50$ and $100$. The data exhibit an excellent linear behavior over the entire range of investigated network volumes, confirming the proposed power-law scaling. Linear regressions yield $\rho(\langle k\rangle)$ for $\langle k\rangle = 10, 20, 50$, and $100$, with $\rho(10) = 0.49(1)$, $\rho(20) = 0.50(1)$, $\rho(50) = 0.49(1)$, and $\rho(100) = 0.48(1)$. The consistency of these values with $\rho \approx 0.50$ indicates that $W_{\varphi}\sim N^{-1/2}$ across the investigated connectivity range. Consequently, the stochastic consensus dynamics becomes increasingly effective at transforming weak initial majorities into predictable collective decisions as the network grows. The observed scaling law provides a quantitative estimate of the reliability gain expected from larger decentralized systems, allowing one to anticipate how much the uncertainty region around the symmetry point $p_c = 0.50$ will contract as additional agents join the consensus network.

\section{\label{sec:discussion}Discussion and outlook}

The stochastic consensus model provides a minimal framework for investigating how local majority interactions and stochastic resolution of locally undecidable configurations drive global consensus in decentralized networks. The numerical results show that the model is highly sensitive to small initial asymmetries around the symmetric point $p_c = 0.50$. For $N = 10^4$ and $\langle k\rangle = 10$, the symmetric case $p = 0.50$ produces positive and negative unanimity with approximately equal probability, while a small bias $p = 0.51$ already leads to positive unanimity in more than $99\%$ of the realizations. For $p\geq 0.52$, all sampled realizations reach positive unanimity. This demonstrates that the SCM amplifies weak initial majorities into robust unanimous outcomes.

Connectivity also plays a central role in the efficiency and reliability of consensus formation. For $N = 10^4$ and $p = 0.51$, increasing the average connectivity from $\langle k\rangle = 5$ to $\langle k\rangle = 1000$ reduces the mean unanimity time from $\langle\tau\rangle \simeq 3 \times 10^5$ to $\langle\tau\rangle \simeq 9 \times 10^4$ UTS. Over the same interval, the positive-unanimity fraction increases from $85.63\%$ to $100\%$, while negative unanimity disappears for sufficiently connected networks. In particular, for $\langle k\rangle\geq 40$, all sampled realizations converge to positive unanimity. These results indicate that increasing connectivity improves both the speed and reliability of consensus formation by enhancing the propagation of local majority information across the network.

The system-size analysis reveals a complementary effect. For fixed $\langle k\rangle = 10$ and $p = 0.51$, the mean unanimity time increases approximately linearly with $N$, indicating that larger networks require more update attempts to reach global agreement. However, larger systems are also more reliable at selecting the initially dominant state. The positive-unanimity fraction increases from $60.31\%$ for $N = 100$ to essentially $100\%$ for large systems, showing that finite-size fluctuations are progressively suppressed as the number of agents grows. This behavior is consistent with the finite-size narrowing of the consensus uncertainty region around $p_c = 0.50$.

The strong amplification of small initial majorities also suggests a potential resilience mechanism in decentralized decision systems. In the SCM, the final consensus state is determined primarily by the sign of the initial majority, and the probability of selecting the majority state rapidly approaches unity as the system volume increases. Consequently, minority perturbations become progressively less effective at altering the final collective outcome. Although the present work does not explicitly model adversarial agents or Byzantine behavior, the observed dynamics indicate that consensus formation is robust to substantial levels of disagreement, provided that a majority state remains. Future investigations incorporating malicious participants, strategic attacks, and heterogeneous trust structures could clarify the extent to which these majority-amplification mechanisms contribute to the security and resilience of practical decentralized systems.

The consensus uncertainty $\varphi=4f_{+\mathbb{1}}f_{-\mathbb{1}}$ reaches its maximum at the symmetry point $p_c = 0.50$, where both absorbing states are equally likely, and vanishes when one final consensus state dominates. By defining $W_{\varphi}$ as the full width at half maximum of $\langle\varphi\rangle$, we introduce the scaling relation $W_{\varphi}\sim N^{-\rho}$, with $\rho \approx 0.50$ for all values of $\langle k \rangle$ investigated. Thus, the region of ambiguous consensus of the SCM outcomes shrinks approximately as $N^{-1/2}$, providing a practical reliability estimate: increasing the number of agents by two orders of magnitude reduces the uncertainty region by approximately one order of magnitude. Therefore, although larger systems require longer times to reach unanimity, they also require progressively smaller initial asymmetries to select a probable consensus outcome.

These findings suggest that the SCM may be useful for decentralized decision systems in which reliability depends on the ability to suppress faulty, adversarial, or inconsistent minority signals. In decision networks, the model captures an abstract mechanism by which a trusted majority can propagate through local interactions and stabilize a global decision. In artificial intelligence workflows that rely on distributed verification, a similar logic could be applied to redundant decision architectures, in which multiple AI agents or model replicas evaluate the same query or decision task. If their outputs can be mapped into discrete agreement states, then compromised or anomalous nodes would appear as minority signals. Provided that the trustworthy fraction remains above the effective majority threshold, the SCM dynamics could act as a decentralized validation layer, reinforcing coherent outputs while suppressing isolated, inconsistent responses.

We emphasize that the SCM is fundamentally a probabilistic consensus mechanism. The quantities $f_{+\mathbb{1}}$, $f_{-\mathbb{1}}$, and $u$ measure the likelihood of different consensus outcomes rather than deterministic guarantees. Therefore, results such as $u=100\%$ indicate that all sampled realizations converged to the same outcome within the observation window used here, though slightly different outcomes are still possible by chance. In practical decentralized systems, this interpretation is natural: reliability is quantified through the probability of reaching the desired consensus state, and the SCM provides a framework for estimating how that probability evolves with network size, connectivity, and initial conditions.

Several limitations remain. The present work considers binary states, random networks, and static average connectivity. Real decentralized systems may involve heterogeneous node capacities, weighted trust, delayed communication, adversarial adaptation, multiple decision states, and evolving network topology. Future work should therefore investigate the SCM on scale-free, small-world, geometric, multiplex, and temporal networks, as well as extensions incorporating reputation, weighted influence, Byzantine agents, and multi-state decisions. These directions may clarify how the consensus amplification and uncertainty-suppression mechanisms observed in this investigation persist in more realistic decentralized systems.\\

\section*{Acknowledgements} \label{sec:acknowledgements}
    The authors acknowledge financial support from Brazilian institutions and funding agents UPE, FACEPE (APQ-1129-1.05/24), CAPES, and CNPq (306336/2025-1). EC gratefully acknowledges support from a SUNY Polytechnic Institute seed grant that funded the creation of CESSAIR, the Center for Safe and Secure AI Robotics, and his work in this project. We used OpenAI's ChatGPT to assist with language refinement and manuscript editing. All scientific content, analysis, and interpretation were solely developed by the authors.


\bibliography{referencesblock}

@misc{Nakamoto2008,
  author       = {Nakamoto, Satoshi},
  title        = {Bitcoin: A Peer-to-Peer Electronic Cash System},
  year         = {2008},
  howpublished = {\url{https://bitcoin.org/bitcoin.pdf}},
  note         = {Online}
}

@article{Castellano2009,
  author  = {Castellano, Claudio and Fortunato, Santo and Loreto, Vittorio},
  title   = {Statistical physics of social dynamics},
  journal = {Reviews of Modern Physics},
  volume  = {81},
  pages   = {591--646},
  year    = {2009}
}

@article{Fontanari2012,
  author  = {Tilles, Paulo F. C. and Fontanari, Jose F.},
  title   = {Mean-field analysis of the majority-vote model broken-ergodicity steady state},
  journal = {Journal of Statistical Mechanics: Theory and Experiment},
  year    = {2012}
}

@article{Peck2017,
  author  = {Peck, Morgan E.},
  title   = {Blockchains: How They Work and Why They'll Change the World},
  journal = {IEEE Spectrum},
  year    = {2017}
}

@article{galam2008,
  author  = {Galam, Serge},
  title   = {Sociophysics: A review of Galam models},
  journal = {International Journal of Modern Physics C},
  volume  = {19},
  number  = {3},
  pages   = {409--440},
  year    = {2008}
}

@article{katarzyna2005,
  author  = {Sznajd-Weron, Katarzyna},
  title   = {Sznajd model and its applications},
  journal = {Acta Physica Polonica B},
  volume  = {36},
  number  = {8},
  pages   = {2537--2547},
  year    = {2005}
}

@article{Steve2016,
  author  = {Huckle, Steve and Bhattacharya, Rituparna and White, Martin and Beloff, Nick},
  title   = {Internet of Things, Blockchain and Shared Economy Applications},
  journal = {Procedia Computer Science},
  volume  = {98},
  pages   = {461--466},
  year    = {2016}
}

@article{Denis2018,
  author  = {Miller, Dennis},
  title   = {Blockchain and the Internet of Things in the Industrial Sector},
  journal = {IT Professional},
  volume  = {20},
  number  = {3},
  pages   = {15--18},
  year    = {2018}
}

@article{Tiago2018,
  author  = {Fernández-Caramés, Tiago M. and Fraga-Lamas, Paula},
  title   = {A Review on the Use of Blockchain for the Internet of Things},
  journal = {IEEE Access},
  volume  = {6},
  pages   = {32979--33001},
  year    = {2018}
}

@article{Thomas2019,
  author  = {McGhin, Thomas and Choo, Kim-Kwang Raymond and Liu, Chien-Ming and He, Debiao},
  title   = {Blockchain in healthcare applications: Research challenges and opportunities},
  journal = {Journal of Network and Computer Applications},
  volume  = {135},
  pages   = {62--75},
  year    = {2019}
}

@article{Mishra2021,
  author  = {Mishra, Lokesh and Kaushik, Vaibhav},
  title   = {Application of blockchain in dealing with sustainability issues and challenges of financial sector},
  journal = {Journal of Sustainable Finance \& Investment},
  volume  = {13},
  number  = {3},
  pages   = {1318--1333},
  year    = {2021}
}

@article{Hisham2023,
  author  = {Mbaidin, Hisham O. and Alsmairat, Mohammad A. K. and Al-Adaileh, Raed},
  title   = {Blockchain adoption for sustainable development in developing countries: Challenges and opportunities in the banking sector},
  journal = {International Journal of Information Management Data Insights},
  volume  = {3},
  number  = {2},
  pages   = {100199},
  year    = {2023}
}

@article{Chen2022,
  author  = {Chen, Yong and Lu, Yang and Bulysheva, Larisa and Kataev, Mikhail Yu.},
  title   = {Applications of Blockchain in Industry 4.0: A Review},
  journal = {Information Systems Frontiers},
  volume  = {26},
  pages   = {1715--1729},
  year    = {2024}
}

@article{Dragoni2020,
  author  = {Herskind, Lasse and Katsikouli, Panagiota and Dragoni, Nicola},
  title   = {Privacy and Cryptocurrencies -- A Systematic Literature Review},
  journal = {IEEE Access},
  volume  = {8},
  pages   = {54044--54059},
  year    = {2020}
}

@article{Allende2023,
  author  = {Allende, Matías and León, Daniel L. and Cerón, Sebastián and others},
  title   = {Quantum-resistance in blockchain networks},
  journal = {Scientific Reports},
  volume  = {13},
  pages   = {5664},
  year    = {2023}
}

@article{Parida2023,
  author  = {Parida, N. K. and Jatoth, C. and Reddy, V. D. and Hussain, Md. M. and Faizi, J.},
  title   = {Post-quantum distributed ledger technology: a systematic survey},
  journal = {Scientific Reports},
  volume  = {13},
  pages   = {20729},
  year    = {2023}
}

@article{Li2020,
  author  = {Li, Xiaofan and Whinston, Andrew B.},
  title   = {Analyzing cryptocurrencies},
  journal = {Information Systems Frontiers},
  volume  = {22},
  pages   = {17--22},
  year    = {2020}
}

@article{Gharavi2024,
  author  = {Gharavi, Hamid and Granjal, Jorge and Monteiro, Edmundo},
  title   = {Post-Quantum Blockchain Security for the Internet of Things: Survey and Research Directions},
  journal = {IEEE Communications Surveys \& Tutorials},
  volume  = {26},
  number  = {3},
  pages   = {1748--1774},
  year    = {2024}
}

@article{Khacef2024,
  author  = {Khacef, Kahina and Benbernou, Salima and Ouziri, Mourad and Younas, Muhammad},
  title   = {A Dynamic Sharding Model Aware Security and Scalability in Blockchain},
  journal = {Information Systems Frontiers},
  volume  = {26},
  pages   = {2323--2336},
  year    = {2024}
}

@article{Sinai2024,
  author  = {Sinai, N. K. and In, H. P.},
  title   = {Performance evaluation of a quantum-resistant blockchain: a comparative study with Secp256k1 and Schnorr},
  journal = {Quantum Information Processing},
  volume  = {23},
  pages   = {99},
  year    = {2024}
}

@article{Alt2025,
  author  = {Alt, Rainer and Gräser, Max},
  title   = {Distributed ledger technology},
  journal = {Electronic Markets},
  volume  = {35},
  pages   = {53},
  year    = {2025}
}

@article{Musilek2021,
  author  = {Lashkari, B. and Musilek, P.},
  title   = {A Comprehensive Review of Blockchain Consensus Mechanisms},
  journal = {IEEE Access},
  volume  = {9},
  year    = {2021}
}

@article{Piva2022,
  author  = {Piva, G. G. and Ribeiro, F. L. and da Mata, A. S.},
  title   = {Voter Model Dynamics on Networks with Social Features},
  journal = {Brazilian Journal of Physics},
  volume  = {52},
  year    = {2022}
}

@article{Dwork1988,
  author  = {Dwork, Cynthia and Lynch, Nancy and Stockmeyer, Larry},
  title   = {Consensus in the presence of partial synchrony},
  journal = {Journal of the ACM},
  volume  = {35},
  number  = {2},
  pages   = {288--323},
  year    = {1988}
}

@article{de1992isotropic,
  author  = {de Oliveira, M. J.},
  title   = {Isotropic Majority-Vote Model on a Square Lattice},
  journal = {Journal of Statistical Physics},
  volume  = {66},
  number  = {1},
  year    = {1992}
}

@article{Ma2017,
  author  = {Ma, Cui-Qin and Qin, Zheng-Yan and Zhao, Yun-Bo},
  title   = {Bipartite consensus of integrator multi-agent systems with measurement noise},
  journal = {IET Control Theory \& Applications},
  volume  = {11},
  number  = {18},
  pages   = {3313--3320},
  year    = {2017}
}

@article{Xie2021,
  author  = {Xie, Guangqiang and Chen, Junyu and Li, Yang},
  title   = {Hybrid-order network consensus for distributed multi-agent systems},
  journal = {Journal of Artificial Intelligence Research},
  volume  = {70},
  pages   = {389--407},
  year    = {2021}
}

@article{baronchelli2025,
  author  = {Ashery, A. F. and Aiello, L. M. and Baronchelli, A.},
  title   = {Emergent Social Conventions and Collective Bias in LLM Populations},
  journal = {Science Advances},
  volume  = {11},
  year    = {2025}
}

@article{Robertson2012,
  author  = {Robertson, Peter J. and Choi, Taehyon},
  title   = {Deliberation, Consensus, and Stakeholder Satisfaction},
  journal = {Public Management Review},
  volume  = {14},
  number  = {1},
  pages   = {83--103},
  year    = {2012}
}

@article{Kurths2021,
  author  = {Wang, W. and Wang, C. and Wang, Z. and Han, B. and He, C. and Cheng, J. and Luo, Xiong and Yuan, Manman and Kurths, J{\"u}rgen},
  title   = {Nonlinear Consensus-Based Autonomous Vehicle Platoon Control Under Event-Triggered Strategy in the Presence of Time Delays},
  journal = {Applied Mathematics and Computation},
  volume  = {404},
  pages   = {126246},
  year    = {2021}
}

@article{granha2022opinion,
  author  = {Granha, Mateus F. B. and Vilela, Andr{\'e} L. M. and Wang, Chao and Nelson, Kenric P. and Stanley, H. E.},
  title   = {Opinion Dynamics in Financial Markets via Random Networks},
  journal = {Proceedings of the National Academy of Sciences},
  volume  = {119},
  number  = {49},
  year    = {2022}
}

@article{Boasson2023,
  author    = {de Jong, G. and Veijer, J.},
  title     = {Cooperative Behavior in Strategic Decision Making: Human Capital and Personality Traits},
  booktitle = {Behavioral Strategy: Emerging Perspectives},
  year      = {2014}
}

@article{Shang2017,
  author  = {Shang, Y.},
  title   = {Consensus in Averager-Copier-Voter Networks of Moving Dynamical Agents},
  journal = {Chaos},
  volume  = {27},
  number  = {2},
  year    = {2017}
}

@article{Mattos2020,
  author  = {de Oliveira, M. T. and Reis, L. H. A. and Medeiros, D. S. V. and Carrano, R. C. and Olabarriaga, S. D. and Mattos, D. M. F.},
  title   = {Blockchain Reputation-Based Consensus: A Scalable and Resilient Mechanism for Distributed Mistrusting Applications},
  journal = {Computer Networks},
  volume  = {179},
  year    = {2020}
}

@article{Baronchelli2018,
  author  = {Baronchelli, A.},
  title   = {The Emergence of Consensus: A Primer},
  journal = {Royal Society Open Science},
  volume  = {5},
  number  = {2},
  year    = {2018}
}

@article{santos1995anisotropic,
  author  = {Santos, M. A. and Teixeira, S.},
  title   = {Anisotropic Voter Model},
  journal = {Journal of Statistical Physics},
  volume  = {78},
  number  = {3--4},
  year    = {1995}
}

@article{vilela2018effect,
  author  = {Vilela, Andr{\'e} L. M. and Stanley, H. E.},
  title   = {Effect of Strong Opinions on the Dynamics of the Majority-Vote Model},
  journal = {Scientific Reports},
  volume  = {8},
  number  = {1},
  year    = {2018}
}

@article{pereira2005majority,
  author  = {Pereira, L. F. C. and Brady Moreira, F. G.},
  title   = {Majority-Vote Model on Random Graphs},
  journal = {Physical Review E},
  volume  = {71},
  number  = {1},
  year    = {2005}
}

@article{campos2003small,
  author  = {Campos, P. R. A. and de Oliveira, V. M. and Brady Moreira, F. G.},
  title   = {Small-World Effects in the Majority-Vote Model},
  journal = {Physical Review E},
  volume  = {67},
  number  = {2},
  year    = {2003}
}

@article{Fadda2022,
  author  = {Fadda, E. and He, J. and Tessone, C. J. and Barucca, P.},
  title   = {Consensus Formation on Heterogeneous Networks},
  journal = {EPJ Data Science},
  volume  = {11},
  pages   = {34},
  year    = {2022}
}

@misc{Schwartz2018,
  author       = {Schwartz, D. and Youngs, N. and Britto, A.},
  title        = {The Ripple Protocol Consensus Algorithm},
  year         = {2018},
  howpublished = {\url{https://ripple.com}}
}

@article{Goles2023,
  author  = {Goles, E. and Medina, P. and Santiv{\'a}{\~n}ez, J.},
  title   = {Majority Networks and Local Consensus Algorithm},
  journal = {Scientific Reports},
  volume  = {13},
  pages   = {1858},
  year    = {2023}
}

@article{Jayabalasamy2024,
  author  = {Jayabalasamy, G. and Pujol, C. and Latha Bhaskaran, K.},
  title   = {Application of Graph Theory for Blockchain Technologies},
  journal = {Mathematics},
  volume  = {12},
  number  = {8},
  pages   = {1133},
  year    = {2024}
}

@article{Anagnostakis2025,
  author  = {Anagnostakis, A. G. and Glavas, E.},
  title   = {Entropy and Stability in Blockchain Consensus Dynamics},
  journal = {Information},
  volume  = {16},
  number  = {2},
  pages   = {138},
  year    = {2025}
}

@article{erdos1960,
  author  = {Erd{\H{o}}s, P. and R{\'e}nyi, A.},
  title   = {On the Evolution of Random Graphs},
  journal = {Publication of the Mathematical Institute of the Hungarian Academy of Sciences},
  volume  = {5},
  pages   = {17--61},
  year    = {1960}
}

@book{bollobas2001,
  author    = {Bollob{\'a}s, B.},
  title     = {Random Graphs},
  series    = {Cambridge Studies in Advanced Mathematics},
  volume    = {73},
  edition   = {2},
  publisher = {Cambridge University Press},
  address   = {Cambridge},
  year      = {2001}
}

@article{dall2002,
  author  = {Dall, J. and Christensen, M.},
  title   = {Random Geometric Graphs},
  journal = {Physical Review E},
  volume  = {66},
  pages   = {016121},
  year    = {2002}
}

@book{Barabasi2016,
  author    = {Barab{\'a}si, A.-L. and P{\'o}sfai, M.},
  title     = {Network Science},
  publisher = {Cambridge University Press},
  year      = {2016}
}

@article{Binder1981,
  author  = {Binder, K.},
  title   = {Finite Size Scaling Analysis of Ising Model Block Distribution Functions},
  journal = {Zeitschrift f{\"u}r Physik B Condensed Matter},
  volume  = {43},
  year    = {1981}
}
\end{document}